\documentclass[a4paper,twoside]{article}

\usepackage{epsfig}
\usepackage{subfigure}
\usepackage{calc}
\usepackage{amssymb}
\usepackage{amstext}
\usepackage{amsmath}
\usepackage{url}
\usepackage{amsthm}
\usepackage{multicol}
\usepackage{pslatex}
\usepackage{apalike}
\usepackage{SCITEPRESS}     

\begin{document}

\title{Managing Usability and Reliability Aspects in Cloud Computing
}

\author{\authorname{Maria Spichkova\sup{1}, Heinz W. Schmidt\sup{2}, Ian E. Thomas\sup{2}, Iman I. Yusuf\sup{2}\\  
Steve Androulakis\sup{3} and Grischa  R. Meyer\sup{3}}
\affiliation{\sup{1}School of Science, RMIT University, 414-418 Swanston Street, 3001, Melbourne, Australia}
\affiliation{\sup{2}eResearch, RMIT University,  	17-23 Lygon Street, 3053, Carlton, Australia}
\affiliation{\sup{3}eResearch, Monash University, Wellington Rd, 3800, Clayton, Australia}
\email{\{maria.spichkova, heinz.schmidt, ian.edward.thomas, iman.yusuf\}@rmit.edu.au,\\
 \{steve.androulakis, grischa.meyer\}@monash.edu}
}

\keywords{Usability, Reliability, Cloud Computing, Modelling, Visualisation 
}

\abstract{ 
Cloud computing provides a great opportunity for scientists, 
as it enables large-scale experiments that cannot are too long to run on local desktop machines. 
Cloud-based computations can be highly parallel, long running and data-intensive, which is desirable for many kinds of scientific experiments. 
However, to unlock this power, we need a user-friendly interface and an easy-to-use methodology for conducting these experiments.
For this reason, we introduce here a formal model of a cloud-based platform and the corresponding open-source implementation. 
The proposed solution allows to conduct experiments  without having a deep technical understanding of cloud-computing, HPC, fault tolerance, or data management in order to leverage the benefits of cloud computing. 
In the current version, we have focused on biophysics and structural chemistry experiments, based on the analysis  of big data from 
synchrotrons and atomic force microscopy. 
The domain  experts noted the time savings for computing and data management, as well as user-friendly interface. \footnote{%
Preprint. Accepted to the 11th International Conference on Evaluation of Novel Approaches to Software Engineering (ENASE 2016). Final version published by SCITEPRESS, 
\url{http://www.scitepress.org}}
}

\onecolumn \maketitle \normalsize \vfill

\section{\uppercase{Introduction}}
\label{sec:intro}

\noindent 
Scientific experiments can be very challenging from a domain point of view, even in the case the computation can be done on a local desktop machine. 
Instruments such as synchrotrons and atomic force microscopy produce massive amounts of data.
Methods used by scientists are frequently implemented in software.
The combination of big data and complex computational methods inevitably requires long running computations and demand for high-performance computing  (HPC) using cluster or cloud computing~\cite{Armbrust2010,Buyya2009}.
HPC and cloud computing marshal large storage resources flexibly and divide tasks and data up to huge numbers of compute cores.
For most users, both present new technologies, either computationally and in data management, and both require learning non-standard data management, programming languages and libraries.

In the case of cloud computing, the users have to learn
how to work within a cloud-based environment, e.g.,  
 how to create and set up virtual machines (VMs), how to collect  the results of their experiments, and finally destroy the VMs, etc.  
Thus, to unlock all the  capabilities  of cloud computing the science users  have to obtain a new set of skills (e.g., knowledge of fault tolerance), 
which might distract from  focusing on the domain specific problems of the experiment.   
 
Cloud computing provides many benefits, e.g., access to online storage and computing   resources  at a moment's notice. 
Nevertheless,  failure while setting up a cloud-based execution environment or  during the execution itself is arguably inevitable:
some or all of the requested VMs may not be  successfully  created/instantiated, or 
the communication with an existing VM may fail due to long-distance network failure -- given clouds data centres are typically remote and communication crosses many network boundaries. Also, one has to realise that all tasks of such parallel computations are required to complete, therefore the failure of any one of them may corrupt the result in some way. 
Statistically this means that the reliability of the overall task completion is the product of that of the individual tasks -- and with very many thousands or millions of compute tasks this may quickly become a vanishable number.

For these  reasons, we propose a user-friendly open-source  platform that would hide the above problems from the user by incapsulating them in the platform's functionality. The feasibility of our platform is shown using case studies across Theoretical Chemical and Quantum Physics group at the RMIT university. 
This paper presents a formal model of a cloud-based platform for scalable and fault-tolerant cloud computations    
as well as its implementation. 
We focus on scientific computations, i.e., we assume that the users of the platform would be researchers working in the fields of physics, chemistry, biology, etc. 

Our solution enables researchers to focus on domain-specific problems, and to delegate to the tool to deal with the detail that comes with accessing high-performance and cloud computing infrastructure, and the data management challenges it poses.  
Moreover, the platform implements various fault tolerance strategies 
to prevent the failed execution from causing %
a system-wide failure,
as well as to recover a failed execution.

\emph{Outline:} 
The rest of the paper is structured as follows.  
Section \ref{sec:model}  presents core features of the proposed model and its implementation as  an open-source  cloud-based platform, including the reliability aspects.   
In Section~\ref{sec:usability},   
we discuss the usability features of the platform and how they are reflected in the conducted  case studies.
Section \ref{sec:related} overviews the related work. 
Section~\ref{sec:conclusion} concludes the paper by highlighting the main contributions, and introduces the future work directions.

\section{\uppercase{Model of a Cloud-based Platform}}
\label{sec:model}

\noindent 
The proposed platform provides access to a distributed computing infrastructure. 
On the logical level it is modelled as a dynamically built set of
{\em Smart Connectors} (SCs), 
which handle the provision of  cloud-based  infrastructure. 
SCs vary from each other by the type of computation to be supported and/or the specific  computing infrastructure to be provisioned. 

\begin{figure}[ht!]
\centering
\scalebox{0.39}{
\includegraphics{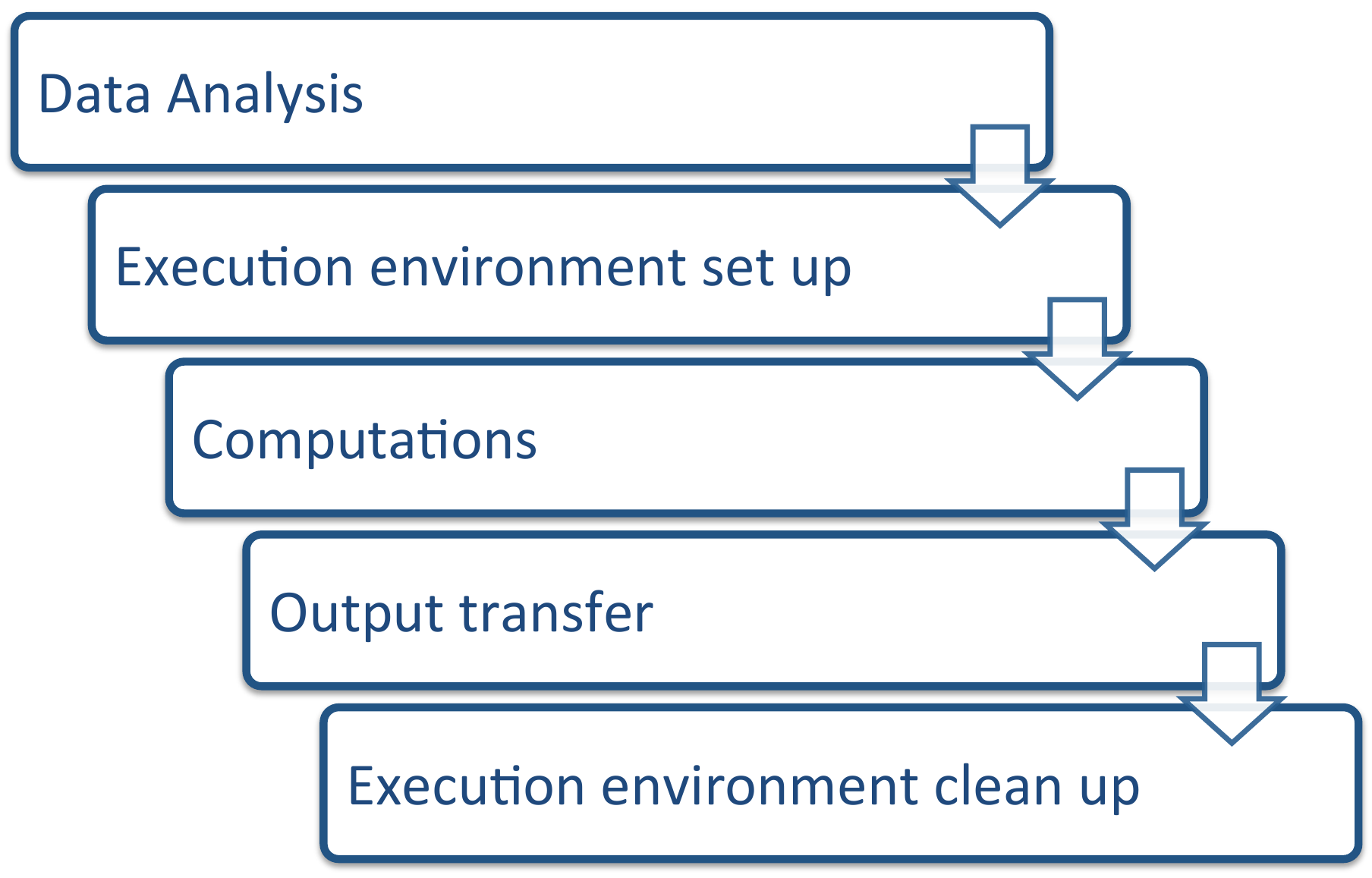}
}
\caption{Cloud service: Workflow}
\label{fig:process1}
\end{figure}

An SC interacts with a cloud service (Infrastructure-as-a-Service) 
on behalf of the user. Figure~\ref{fig:process1} presents the corresponding workflow. 
With respect to the execution environment, the only information that is expected from the user is to specify the number of computing resources she wishes to use, credentials to access those resources, and the location for transferring the output of the computation. 
Thus, the user does not   need to know about   how  the execution environment is set up (i.e., how
  VMs  are created and configured  for the upcoming simulation),  how  a simulation  is executed, how the final output is transferred and how the environment is cleaned up after the computation completion (i.e., how the VMs are destroyed).

 \begin{figure*}[ht!]
\centering
\scalebox{0.5}{
\includegraphics{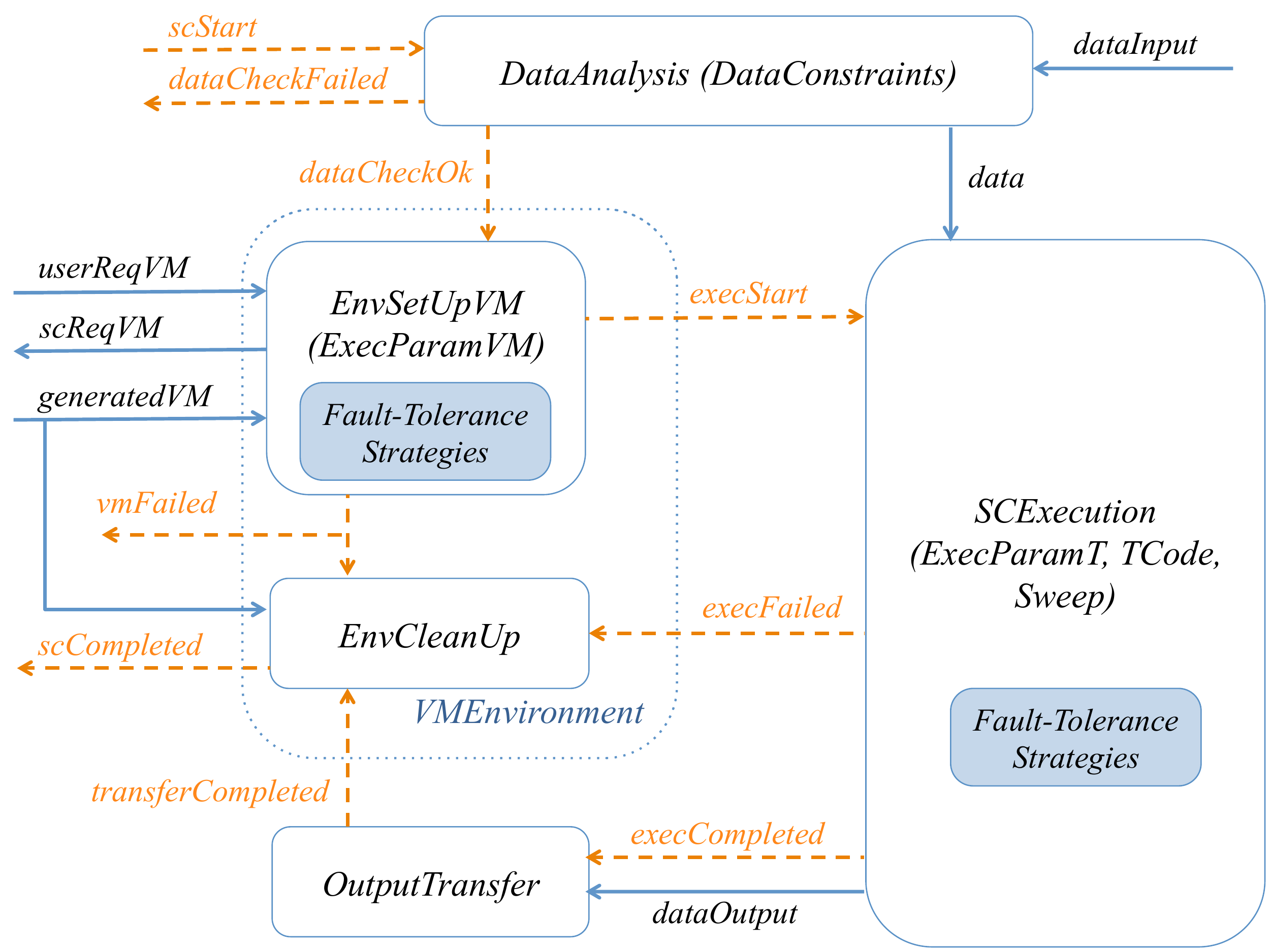}
}
\caption{Logical architecture of a \emph{Smart Connector}.}
\label{fig:chiminey}
\end{figure*}

~\\
Figure~\ref{fig:chiminey} shows the logical architecture of an SC.
Each SC consist of five logical components, which can also be seen as processes within the framework workflow, 
presented on Figure~\ref{fig:process1}.
The model is based on our previous research on process analysis and specification
 \cite{mise2015process,spichkova_processes}. 
 While developing the model we focused on the understandability and readability aspects~\cite{Spichkova2013HFFM,spichkova2013we}. 

We specify for any process $P$ its entry and exit points by $Entry(P)$ and $Exit(P)$ respectively, and
represent a process $P$ (elementary or composed) by the corresponding component specification $PComp$. 
All the control channels (representing entry and exit points of a process) are drawn as orange dashed lines,
the corresponding auxiliary components over these channels are also drawn in orange. 

An execution of an SC is called a \emph{job}.
An SC executes a user requested \emph{process cP}, which consists of tasks $Task_1, \dots, Task_{NT}$, which could be executed in iterative manner. 
In the simplest case, \emph{cP} consists of a single task that should be executed once only.

A concrete SC is build from a general template by configuration its parameters: 
\begin{itemize}
\item
\emph{DataConstraints} specifies constraints on the user provided input \emph{dataInput};
\item
\emph{ExecParamVM} specifies parameters of the job, e.g., which compilers should be installed on the generated VMs;
\item
\emph{ExecParamT} is  a list of the task execution parameters $ExecParamT_1$, \dots, $ExecParamT_{NT}$. 
These parameters specify for each task which data are required for its execution, what is a convergence criterion and whether there is any for that task, which scheduling constrains are required, etc.; 
\item
\emph{TCode} presents an actual executable code for the corresponding tasks, in general case it consists of $NT$ elements.
\item
\emph{Sweep} is a list of values to sweep over: 
With respect to configuring and executing the simulation, the user may set the value and/or ranges of domain specific parameters, and subsequently   automatically creating and executing multiple instances of the given SC to sweep across ranges of values.  
\end{itemize}
The first three parameters can be partially derived from \emph{TCode} on the development stage for a concrete Smart Connector.

~\\
The \emph{DataAnalysis} component is responsible for the preliminary check  whether 
the user \emph{dataInput} satisfies the corresponding \emph{DataConstraints}, both on syntactical and on semantical level.  
The  \emph{DataAnalysis} process is started by receiving an \emph{scStart} signal from the user.
If the data check was successful, the \emph{VMEnv} component is activated by signal \emph{dataCheckOk} 
and the data are forwarded to the \emph{SCExecution} component, otherwise the process is stopped and the user receives an error message \emph{dataCheckFail}. 
 
 ~\\
The \emph{EnvSetUpVM}  component is responsible for the communication with the cloud to obtain a number of VMs 
that is enough for the task according to the user requests \emph{userReqVM}.
The user request \emph{userReqVM} is a pair of numbers $(iN,mN)$, where
$iN$ is an ideal and $mN$ is a minimal (from the user's point of view) number of VMs required for the experiment. 
The \emph{EnvSetUpVM} component requests from the cloud $iN$ VMs. 

~\\ 
\textbf{Fault-Tolerance properties of  \emph{EnvSetUpVM}:} \\
If some of the requested VMs are not  created/instantiated successfully 
(i.e. only $j$ VMs are successfully created, where $0 \le j < mN$. 
The SC will employ various strategies to create the remaining VMs: 
it will retry to make either  a block  request to create $mN - j$ VMs  at once or  a single request at a time. 
For these purposes, one of the parameters within \emph{ExecParamVM} have to be  \emph{RetryLimit}, which limits the number of retries. 

 \begin{figure*}[ht!]
\centering
\scalebox{0.5}{
\includegraphics{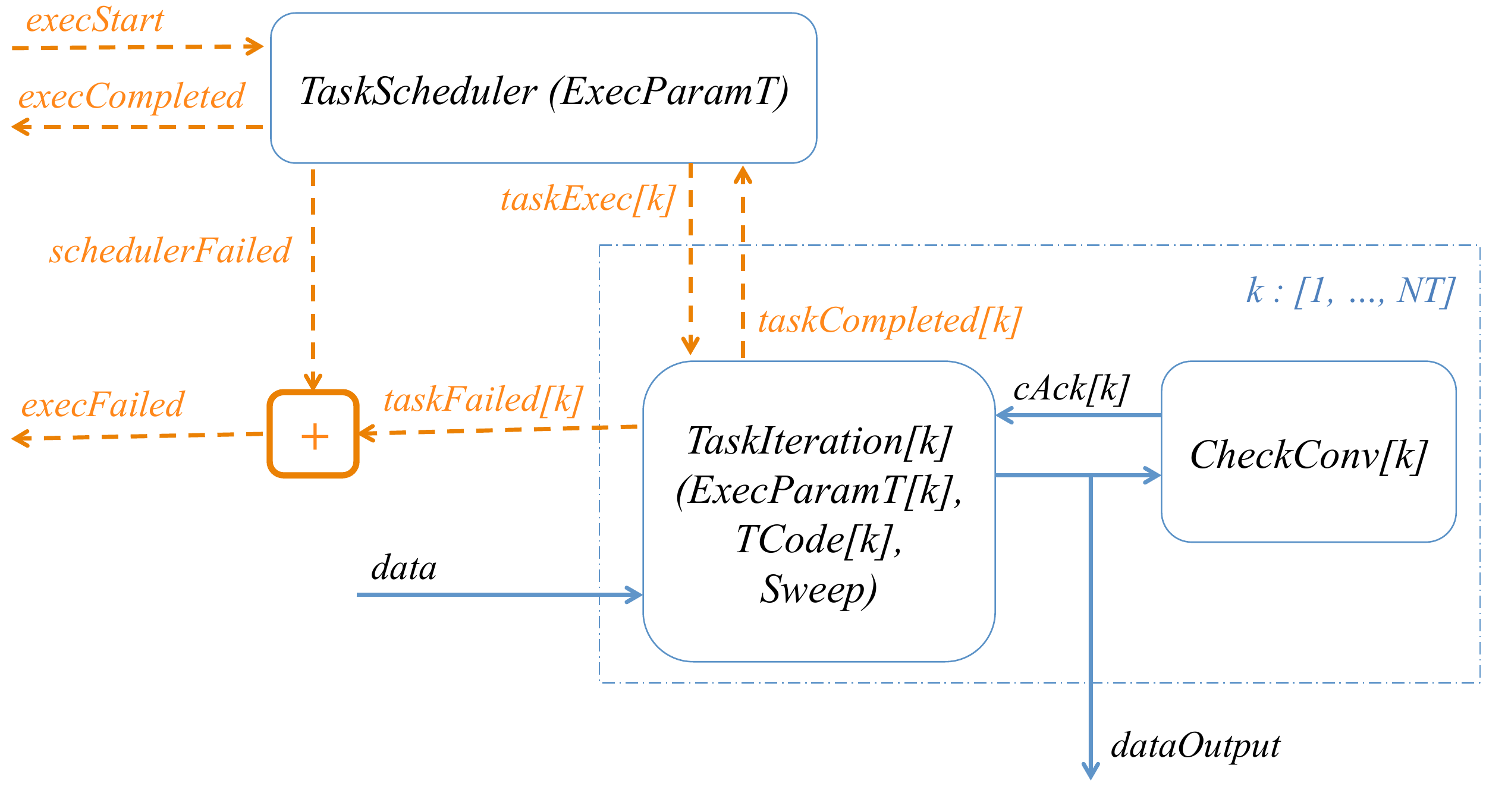}
}
\caption{\emph{SCExecution} subcomponent of Smart Connector.}
\label{fig:scexecution}
\end{figure*}

The number of VMs generated by cloud (the list \emph{generatedVM} represents the list of the corresponding identifiers, e.g., IP addresses) has to fulfil the following property
\begin{equation}
\label{eq:eq1}
mN \le \emph{length(generatedVM)}  \le iN
\end{equation}
If Equation \ref{eq:eq1} is not fulfilled (more precisely, if \emph{length(generatedVM)} $< mN$, because the cloud never provides a number of VMs larger than requested), 
 or the bootstrapping failed (e.g. due lost connection to the cloud) the process is stopped 
 and the user receives an error message \emph{vmFail}. 
 The message \emph{vmFail} is also activates the \emph{EnvCleanUp}  component, 
 which is responsible for final clean up of the system and the destruction of corresponding VMs. 
 When the clean up is completed, the \emph{EnvCleanUp}  component generated the message \emph{scCompleted}, which also indicates that the whole process chain is completed.

 If Equation \ref{eq:eq1} is fulfilled, our platform preforms bootstrapping of generated VMs, and the required compilers are installed according to \emph{ExecParamVM}. 
If the bootstrapping was successful, the \emph{SCExecution} component is activated by the signal \emph{execStart}.  
Then we could have two cases: 
\begin{itemize}
\item
The smart connector execution was successful. Then, the  \emph{SCExecution} component  
	\begin{itemize}
	\item 
	forwards the results of the computations \emph{dataOutput} to the  \emph{OutputTransfer} component;
	\item
	generates the signal \emph{transferStart} that
	  indicates that the \emph{SCExecution} process is completed, and 
	  activates  \emph{OutputTransfer}.  
	\end{itemize}
The \emph{OutputTransfer} component is responsible for the transfer of the output data to the corresponding server
and to a  
data management system.
When the data transfer is completed,  
 the message \emph{transferCompleted} is generated 
to activate the \emph{Env\-CleanUp}  component.
\item
The smart connector execution failed on the stage of scheduling or during execution of a task. 
Then, the  \emph{SCExecution} component generates the signal \emph{execFailed}, to activate  \emph{EnvCleanUp}  
 for the final clean up of the system and the destruction of corresponding VMs.  
\end{itemize}

~\\
 \emph{EnvSetUpVM} and \emph{EnvCleanUp} 
can also be logically composed into a meta-component 
\emph{VMEnv}, which is responsible for any communication with the cloud and the corresponding environment manipulations.

The \emph{SCExecution} component (cf. Figure~\ref{fig:scexecution}) is the main part of a smart connector.   
It  is responsible for the actual execution of the task and  provides a number of the task execution options,  
defined by parameters \emph{ExecParamT}.

~\\
\textbf{Fault-Tolerance properties of  \emph{SCExecution}:}\\ 
The computation might fail  due to  network or  VM failure, i.e., 
the VM that hosts some of the processes cannot be reached.
To avoid an endless  waiting on  the output from the processes on the unreachable VM,  
 the smart connector will identify the processes that are hosted there, %
and then execute the appropriate fault tolerance strategy, e.g.,
$(i)$ marking the processes that are hosted on the unreachable VM as failed beyond recovery and then collecting the output of   processes from the other VMs, or 
$(ii)$ re-running the failed processes on a different VM until maximum re-run limit is reached.
However, we do not implement any strategies to recover a failed process if the failure was due to an internal bug within the task code.  
In this case, a smart connector will notify the user about the detected failure, as this provides an opportunity to correct the bug.

~\\
\emph{SCExecution} has the following subcomponents: 
\begin{itemize}
\item
 \emph{TaskScheduler}: responsible for scheduling of the tasks and their execution in the right order;
 \item
 \emph{TaskIterarion}: responsible for execution of task iterations 
 according to the corresponding task code; in general case we have $NT$ tasks, where $NT \ge 1$. Thus, a connector has $NT$ components  \emph{TaskIterarion}, one for each task (which means that each task should have at least one iteration of its execution);
\item
\emph{CheckConv} is an optional component, to check whether convergence criterion of a multi-iterational execution is met. 
\end{itemize}

 ~\\
Our model allows us not only to have a precise and 
concise specification of the cloud-based platform on a logical level 
but also provides a basis for a formal analysis of its properties, including security properties, as well as of the core computation properties.
For the formal analysis we suggest to use an interactive semi-automatic theorem prover Isabelle/HOL \cite{npw,SledgehammerSpass} and the corresponding metho\-dologies 
\cite{FocusStreamsCaseStudies-AFP,CryptoBasedCompositionalProperties-AFP,spichkova2014formalisation}, 
as 
the provided specification is compatible to these methodologies. 
Moreover, the purposed representation gives a basis for the resource management and performance prediction, cf. \cite{perfPrediction}, 
as it allows a straightforward analysis of the worst case execution time (WCET) of the composed processes.

  The early results on the platform open-source  implementation were presented in \cite{Chiminey,ChimineyICPADS}. 
 The current version of the platform provides a set of APIs to create new and customise existing SCs. 
We do not restrict our system to be build using a single programming language. Python was chosen as the development language due to its rapid prototyping features, integration with our data curation system, and due to its increasing uptake by researchers as a scientific software development language. However, the domain-specific calculations could be written in any language. The choice of the language depends on the domain and the concrete research task which should be solved.

\section{\uppercase{Usability aspects}}
\label{sec:usability}

\noindent 
The proposed open-source  platform has been applied across two research disciplines, physics (material
 characterisation) and structural biology (understanding materials at the atomic scale), to assess its usability and practicality.   
 The domain  experts noted the following advantages of the platform:
 \begin{itemize}
 \item
  time savings for computing and data management, 
  \item
  user-friendly interface for the computation set up,
  \item
  visualisation of the calculation results as 2D or 3D graphs.
  \end{itemize}

\begin{figure}[ht!]
\centering
\scalebox{0.31}{
\includegraphics{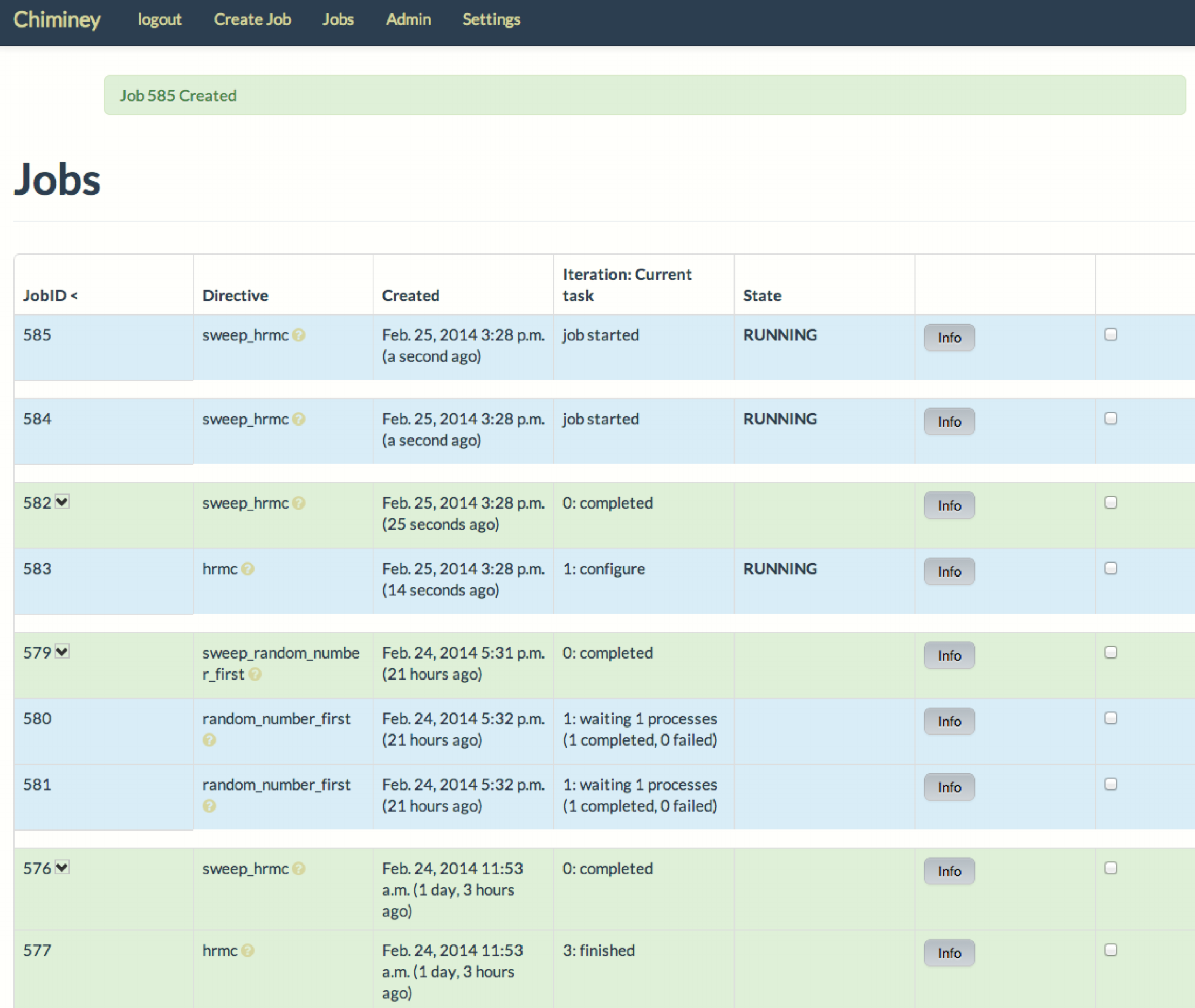}
}
\caption{\emph{Jobs} section of the platform.}
\label{fig:jobs}
\end{figure}

\noindent
The menu has the following sections: \emph{Logout}, \emph{Create Job}, \emph{Jobs}, \emph{Admin}, \emph{Settings}.
 After logging in, the users are in the  \emph{Jobs} section, where they can see the status of current and previous jobs (executions of SCs).  
 Some of the jobs may be processing, have been completed or had errors (cf. Figure \ref{fig:jobs}).

\begin{figure}[ht!]
\centering
\scalebox{0.29}{
\includegraphics{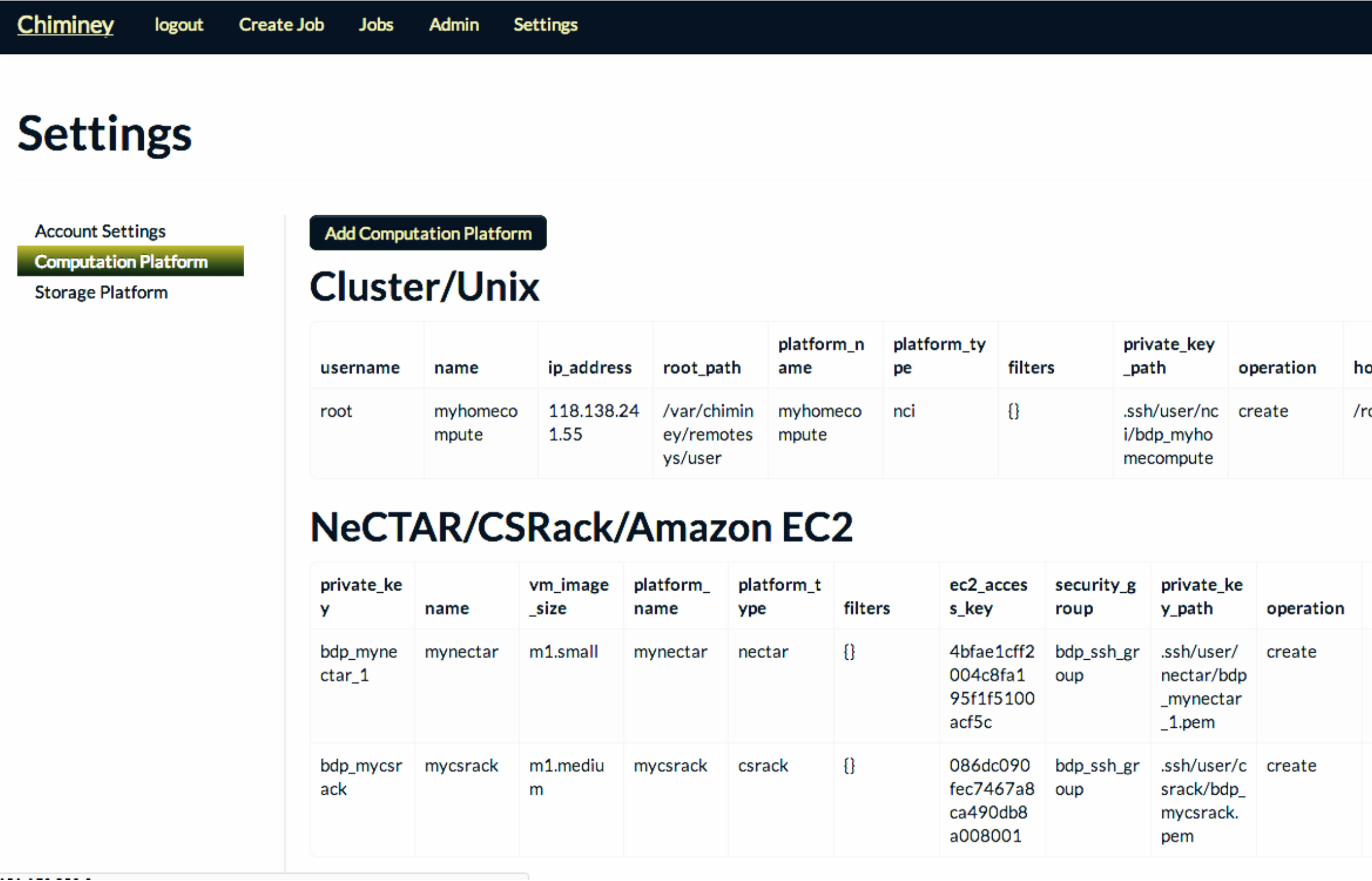}
}
\caption{\emph{Settings} section of the platform}
\label{fig:settings}
\end{figure}

~\\
In the  \emph{Settings} section, we can set up general account properties as well as change settings of computation and storage platforms, cf. Figure \ref{fig:settings}. 

\begin{figure}[ht!]
\scalebox{0.33}{
\includegraphics{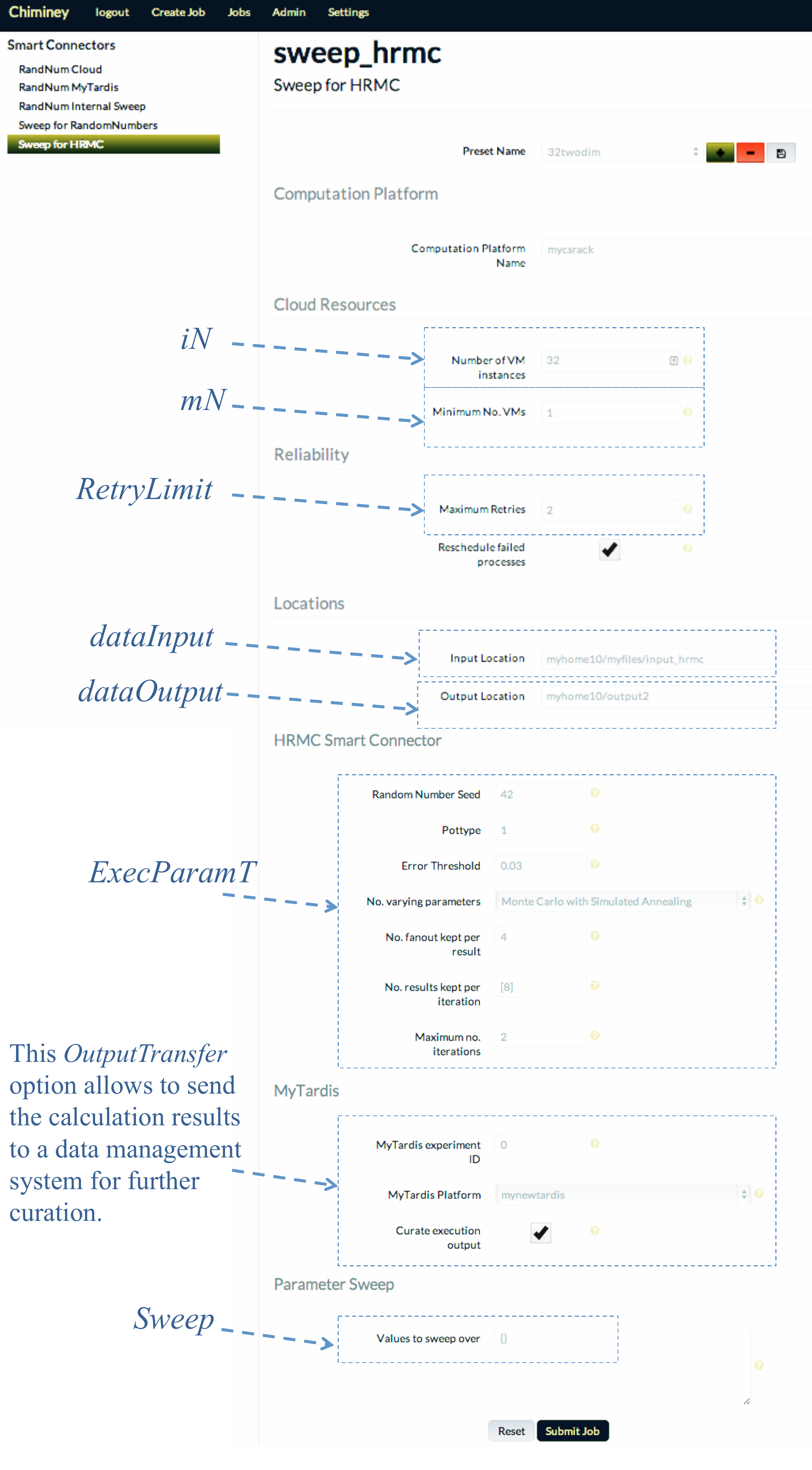}
}
\caption{\emph{Create Job} section of the platform.
} 
\label{fig:jobcr}
\end{figure}

~\\
When we select \emph{Create Job} in the menu,  we will  see the job submission page, which has a set of available SCs currently registered.  
Figure \ref{fig:jobcr} shows how to create a job on example of execution of Monte Carlo simulations, which was a part of one of our case studies (we extended the print screens with the comments to show the match of these parameters to the model from Section~\ref{sec:model}). 
In that case study, the
 Monte Carlo based simulations were applied for modelling of a material's porosity and the size distribution of its pores (industrial applications of these research are in diverse areas such as filtration and gas adsorption).
  One such modelling methodology is the Hybrid Reverse Monte Carlo (HRMC), cf. \cite{Opletal2008}. 
HRMC characterises a material's microstructure by producing models consistent with experimental diffraction data, while at the same time ensuring accurate local bonding environments via energy minimisation. 
 
\begin{figure}[ht!]
\centering
\scalebox{0.28}{
\includegraphics{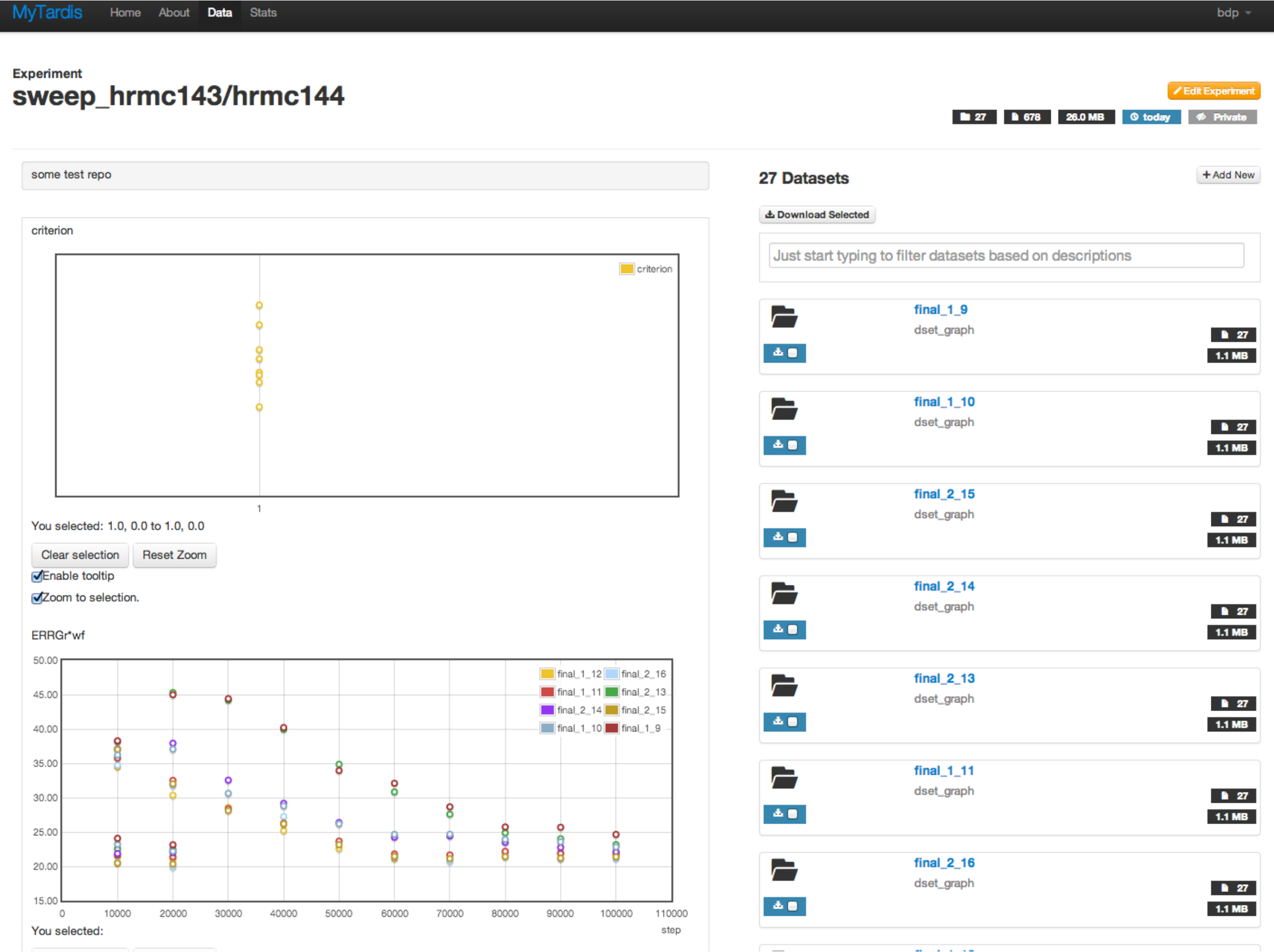}
}
\caption{Visualisation of calculation results.}
\label{fig:gr1}
\end{figure}

\begin{figure}[ht!]
\centering
\scalebox{0.28}{
\includegraphics{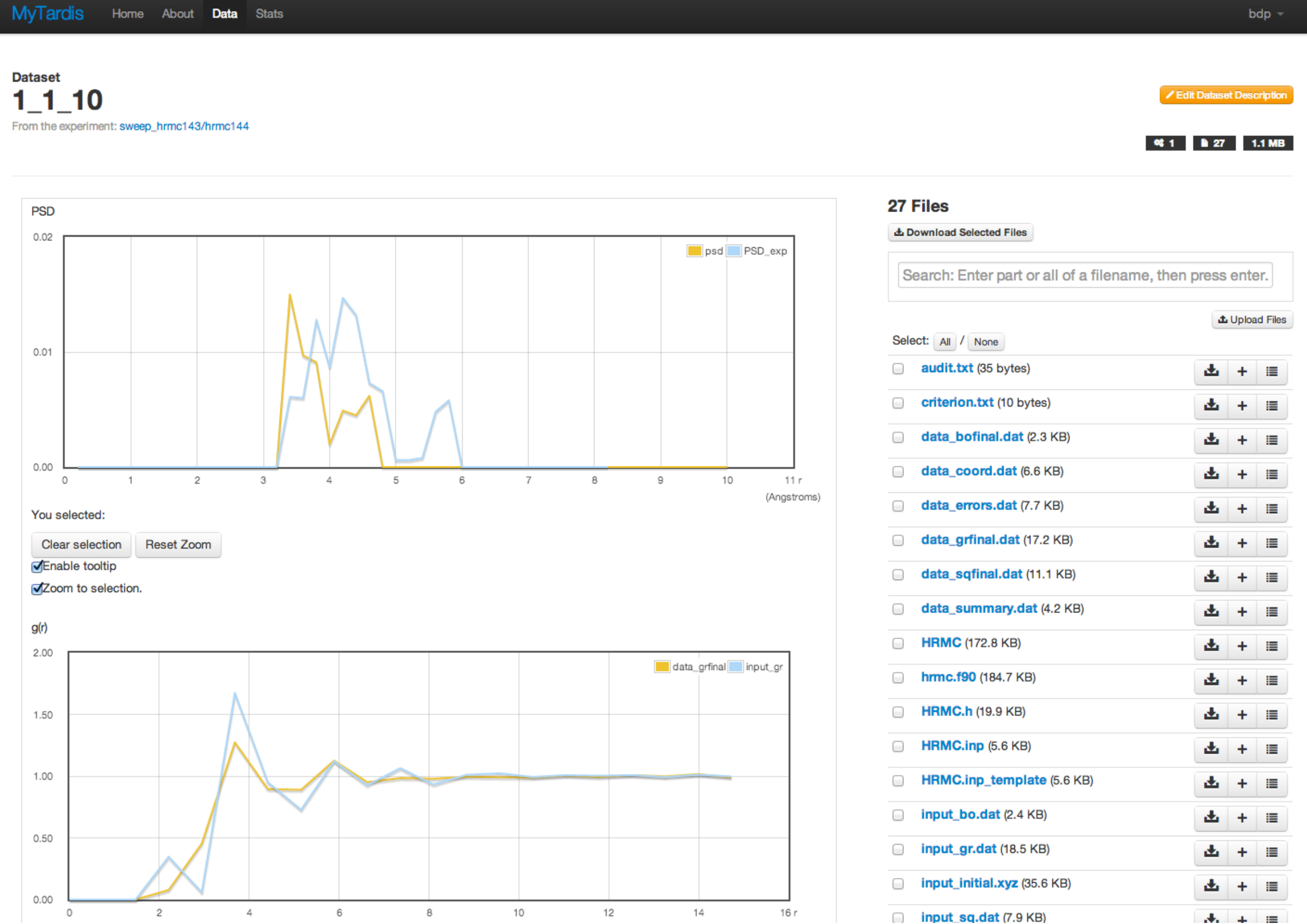}
}
\caption{Visualisation of calculation results as 2D graphs.}
\label{fig:gr2}
\end{figure}

 The user interface, combined with
the MyTardis~\cite{Androulakis2008} data curation module, allows for flexible handling of data 
 according to its completion and significance. 
The results of the calculation can be visualised as 2D or 3D graphs using a plug-in developed to provide better readability of the obtained data (cf. Figures \ref{fig:gr1} and \ref{fig:gr2} for examples). The curated datasets are fully accessible and shareable online.
The generated graphs can easily be used for presentations or in written documents.
\\
 
\newpage

\section{\uppercase{Related Work}}
\label{sec:related}

\noindent 
 While developing the model, we focused on its understandability and readability aspects.  
There are several approaches on model readability, cf. \cite{Wimmer2006,Mendling2007,Zugal2012,spichkova2013we,spichkova2015towards}.

 The development of formal models and architectures for system involved
in cloud computing is a more recent area of system engineering, cf.
\cite{Vaquero2008}, \cite{Buyya:2008:SUO:1395082.1395624}, \cite{Leavitt:2009}, \cite{zhang2010cloud}, \cite{CComp2010}, 
 \cite{GridsCloudsV}.
 
Several approaches have proposed the data stream processing systems
for clouds, e.g., 
\cite{Martinaitis:2009} introduces an approach towards component-based
stream processing in clouds, \cite{DataStreamSharing} presents a
work on data stream sharing.  %
Yusuf and Schmidt  
have shown that the fault-tolerance is best achieved by  reflecting the computational flow in such complex scientific system architectures, cf. 
\cite{Yusuf2013}. 

 There are different types of  scientific workflow systems such as Kepler~\cite{Ludascher2006}, Taverna~\cite{Oinn2006} and Galaxy~\cite{Afgan2011}, which are designed to allow researchers to build their own workflows. 
 The  \emph{contribution of the work presented in this paper} is that our platform provides  drop-in components, Smart Connectors,  for  existing workflow engines:  
$(i)$ researchers can utilise and adapt existing Smart Connectors;
$(ii)$ new types of  Smart Connectors would be developed within the framework if necessary.
From our best knowledge, there is no other framework having this advantage. 
SCs are geared toward providing power and flexibility over simplicity.

Nimrod~\cite{Buyya2000} is a set of software infrastructure for executing large and complex computational, 
contains a simple language for describing sweeps over parameter space and the input and output of data for processing.  
Nimrod is  compatible with the Kepler system ~\cite{Ludascher2006}, 
such that users can set up complex computational workflows and have them executed without having to interface 
directly with a high-performance computing system. 

One of the directions of our future work is incorporation Nimrod into our open-source platform  for the execution of its Smart Connectors. 
However, Nimrod's web API is currently in development, making interfacing with its capabilities non-trivial in a web-based cloud environment.

\section{\uppercase{Conclusions}}
\label{sec:conclusion}

\noindent 
Cloud computing provides a great opportunity for scientists, but to unlock all its benefits, we require a platform with a user-friendly interface and an easy-to-use methodology for conducting the experiments. 
Usability and reliability features are crucial for such systems.
This paper presents a model of a cloud-based platform and the latest version of its open-source implementation, focusing on usability and reliability aspects. 
The proposed platform allows to conduct the experiments  without having a deep technical understanding of cloud-computing, HPC, 
fault tolerance, or data management  in order to leverage the benefits of cloud computing. 

We believe that the proposed platform will have a strong positive impact on the research community, because it give an opportunity to focus on the main research problems and takes upon itself solving of the major part of the infrastructure problems.

~\\
\emph{Future work:} 
The main direction of our future work is application of the platform for an efficient  testing 
 based on analysis of system architecture.

\section*{\uppercase{Acknowledgements}}

The Bioscience Data Platform project  
acknowledges funding from the NeCTAR project No. 2179~\cite{nectar2013}. 

We also would like to thank our colleagues  
Dr~Daniel~W. Drumm (School of Applied Sciences, RMIT University),
Dr~George Opletal (School of Science, RMIT University),
Prof~Salvy~P. Russo  (School of Science, RMIT University), and
Prof~Ashley~M.  Buckle (School of Biomedical Sciences, Monash University) 
for the fruitful collaboration within the Bioscience Data Platform project.

\bibliographystyle{apalike}

\vfill
\end{document}